\documentclass[aps,preprint]{revtex4}%
\usepackage{amsfonts}
\usepackage{amsmath}
\usepackage{amssymb}
\usepackage{graphicx}%
\begin{document}
\title{1D Cahn-Hilliard equation : Ostwald ripening and modulated phase systems}
\author{Simon Villain-Guillot}
\affiliation{Centre de Physique Mol\'{e}culaire Optique et Hertzienne, Universit\'{e}
Bordeaux I, 351 cours de la Lib\'{e}ration 33405 Talence Cedex, France }
\email{s.villain@cpmoh.u-bordeaux1.fr}

\begin{abstract}
Using an approximate analytical solution of the Cahn-Hilliard equation
describing the coalescence during a first order phase transition, we compute
the characteristic time for one step of period doubling in Langer's self
similar scenario for Ostwald ripening. As an application, we compute the
thermodynamically stable period of a 1D modulated phase pattern.

\end{abstract}
\maketitle

Pacs numbers~: 05.45.Yv, 47.20.Ky, 47.54.+r,64.75.+g\bigskip

\section{Introduction}

When a homogenous system departs suddenly from equilibrium, it will
spontaneously segregate into two different states, thermodynamically more stable.

This process can either initiate via a nucleation process, where an energy
barrier has to be crossed, or via a spinodal decomposition when the system is
led into a linearly unstable configuration. In this latter case, the leading
instability selects a modulation of the order parameter at a well defined
length scale. This instability will grows and, due to nonlinearity, saturates.
The resulting micro-segregated pattern is composed of well defined interfaces
delimiting monophasic domains containing one of the two stable phases. These
interfaces will then interact with each other and coalesce, during a much
slower, self-inhibiting process, where the number of domains diminishes
whereas their typical size increases, ending with the formation of either a
micro-segregated complex pattern \cite{SC}, or a single interface separating
two semi infinite domains, one for each new stable phase (macro-segregation).

Hillert\cite{hillert}, Cahn and Hilliard\cite{CH} have proposed a model
equation describing the segregation for a binary mixture. This model, known as
the Cahn-Hilliard equation (C-H later on), belongs to the Model B class in
Hohenberg and Halperin's classification \cite{halperin}. It is a standard
model for phase transition with conserved quantities and has applications to
phase transition in alloys \cite{hillert}, binary mixtures\cite{wagner}, vapor
condensation \cite{beg} and liquid crystals\cite{coullet}, segregation of
granular mixtures in a rotating drum\cite{oyama}, or formation of sand ripples
\cite{melo,stegner}.

In this article, by spinodal decomposition, we refer to the first stage of the
dynamics only, while coarsening or Ostwald ripening will denote the second
stage. Although this coarsening dynamics is in fact already present, its
influence can be neglected during the first stage of the process.

An important activity has been devoted to the description of the dynamics of
phase transition, using experiments, numerical simulations \cite{baron} or
scaling methods \cite{bray}. The late stage of the dynamics, where the Ostwald
ripening dominates, exhibits \textquotedblright dynamical
scaling\textquotedblright~: the dynamics presents a self-similar evolution
where time enters only through a length scale $L(t)$, associated with a
typical length of the domains. This scaling argument gives the law $L(t)\sim
t^{1/3}$ for spatial dimensions greater than one and a logarithmic behavior in D=1.

This last stage, as observed in two-dimensional demixion of
copolymers\cite{copol} and as suggested initially by Langer\cite{langer}, can
be described as a process of synchronous fusion and evaporation of domains,
spatially alternated. It corresponds to the evolution that breaks the least
the symmetries of the pattern. This observation motivated our work and the aim
of this article is to make use of a one dimensional ansatz to describe
quantitatively the ideal coalescence process. This ansatz is in the form of a
one parameter family of symmetric profiles which interpolates between two
stationary states composed of homogeneous domains of length $\lambda$ and
2$\lambda$. It allows in 1D to describe a self similar sequence of coalescence
and evaporation\ of domains (or of their dual counterparts, the interfaces),
starting from the periodic micro-segregated state which ends the spinodal
decomposition dynamics, and leading continuously to either a single interface,
or a stable 1D modulated phase.

The paper is organized as followed: part 2 will focus on the first part of the
dynamics, i.e. spinodal decomposition.
%%%%%%%%%%%
We will reproduce briefly the original derivation by Cahn and Hilliard,
restricting ourselves to the one dimensional case, mainly to fix the notations
(part 2.1).
%%%%%%%%%%%%%%
In part 2.2, a family of symmetric solutions of the Ginzburg-Landau equation
will be used to study the non linear part of dynamics and in part 2.3 we will
identify all the intermediate symmetric stationary states of the (C-H)
dynamics. Then in part 3, we will turn to Ostwald ripening : in part 3.1, a
non-symmetric family of solutions of the (G-L) equation is used to construct a
continuous interpolation between two consecutive symmetric stationary states.
After a study of the energy landscape associated with this ansatz (part 3.2),
we will compute in part 3.3 the characteristic time associated with one step
of coalescence : we will recover the 1D logarithmic law. Before concluding, as
an application in part 4, our ansatz will be used, in the case of
micro-segregation, to compute the period of the final thermodynamically stable
modulated pattern.

\newpage

\section{The Cahn-Hilliard model}

\subsection{Linear stability Analysis}

The Cahn-Hilliard (or Conservative Time Dependant Ginzburg Landau) equation is
a modified diffusion equation for the scalar order parameter $\Psi$, which
reads in its dimensionless form:%

\begin{equation}
\frac{\partial\Psi}{\partial t}\left(  \mathbf{r},t\right)  =\mathbf{\nabla
}^{2}\frac{\delta F_{GL}(\Psi)}{\delta\Psi}=\mathbf{\nabla}^{2}(\frac
{\varepsilon}{2}\Psi+2\Psi^{3}-\mathbf{\nabla}^{2}\Psi). \label{CHeq}%
\end{equation}
The real order parameter can correspond to the dimensionless magnetization in
Ising ferromagnet, to the fluctuation of density of a fluid around its mean
value during a phase separation or to the local concentration of one of the
components of a binary solution. $\varepsilon$ is the dimensionless control
parameter of the system ; it is often identified to the reduced temperature
($\varepsilon=\frac{T-T_{c}}{T_{c}}$ where $T_{c}$ is the critical temperature
of the first order phase transition). This equation, first derived by Cahn and
Hilliard, has also been retrieved by Langer\cite{langer} from microscopic
considerations. A conservative noise can be added to account for thermal
fluctuations \cite{cook}, but in this article, we will only consider the
noiseless (C-H) equation.

It admits homogeneous stationary solutions which are extrema of the symmetric
Landau potential $V(\Psi)=\frac{\varepsilon}{4}\Psi^{2}+\frac{1}{2}\Psi^{4}$.
For positive $\varepsilon$, there is only one homogenous solution $\Psi=0$
which is linearly stable. A pitchfork bifurcation can be experienced when
quenching the system from a positive reduced temperature $\varepsilon$ to a
negative one : the $\Psi=0$ solution becomes unstable ; two other symmetric
stable solutions appear $\Psi=\pm\frac{\sqrt{-\varepsilon}}{2}$.

%%%When the equation is studied for a constant $\varepsilon $, via a rescaling of $\Phi $ (as $\sqrt{-\varepsilon }\Phi $), position $\QTR{bf}{r}$ (as $\QTR{bf}{r}/\sqrt{-\varepsilon }$) and time (as $t/\QTR{group}{\left| \varepsilon \right| }^{2}$), we observe that we could restrict the dynamics to the case $\varepsilon =-1$. However, since we will later on compare stationary solutions of (C-H) with different reduced temperature, we will continue to write the equation with a given $\varepsilon $, keeping in mind that the dynamics can always be rescaled to the case $\varepsilon =-1$.\newline{}
Spinodal decomposition is the dynamics resulting of such a quench. Cahn and
Hilliard have studied the early times of this dynamics by linearizing equation
(\ref{CHeq}) around $\Psi=0$ (i.e. neglecting the non linear term $\Psi^{3}$).
Considering $\Psi$ as a sum of Fourier modes:%

\begin{equation}
\Psi(\mathbf{r},t)=\sum_{\mathbf{q}}\Psi_{\mathbf{q}}e^{i\mathbf{q\cdot
r}+\sigma t}%
\end{equation}
where $\Psi_{\mathbf{q}}$ is the Fourier coefficient at $t=0$, they obtained
for the amplification factor $\sigma(\mathbf{q})$~:%

\begin{equation}
\sigma(\mathbf{q})=-(\mathbf{q}^{2}+\frac{\varepsilon}{2})\mathbf{q}^{2}%
\end{equation}
This shows immediately that $\Psi=0$ is linearly stable for $\varepsilon>0$
while a band of Fourier modes are unstable for negative $\varepsilon$, since
$\sigma(\mathbf{q})>0$ for $0<q<\sqrt{(-\varepsilon/2)}$. Moreover, the most
unstable mode is for $q_{C-H}=\sqrt{-\varepsilon}/2$. This wave number of
maximum amplification factor will dominate the first stage of the dynamics;
this explains in particular why the homogeneous domains appear at length
scales close to $L=\lambda_{C-H}/2=\pi/q_{C-H}$, half the wave length
associated with the instability. For longer times, interfaces separating each
domain interact through Ostwald ripening, causing $<L>$ to change slowly
toward higher values \cite{baron,izu}

We will now use known results on non-homogeneous solutions of the (G-L)
equation to study both the saturation of the spinodal decomposition and the coalescence.

\subsection{Stationary States of the Cahn-Hilliard Dynamics : Symmetric
Soliton Lattice Solutions}

For $\varepsilon<0$, \ there exists a whole family of solution of the one
dimensional (G-L) equation~:%

\begin{equation}
\frac{\varepsilon}{2}\Psi+2\Psi^{3}-\nabla^{2}\Psi=0 \label{CH0}%
\end{equation}
These solutions, the so-called soliton-lattice solutions, are~:
\begin{equation}
\Psi_{k,\varepsilon}(x)=k\Delta\mathrm{Sn}(\frac{x}{\xi},k)\text{ with }%
\xi=\Delta^{-1}=\sqrt{2\frac{k^{2}+1}{-\varepsilon}} \label{amplitude}%
\end{equation}
where $\mathrm{Sn}(x,k)$ is the Jacobian elliptic function sine-amplitude, or
cnoidal mode. This family of solutions is parametrized by $\varepsilon$ and by
the Jacobian modulus $k\in\left[  0,1\right]  $, or \textquotedblright
segregation parameter\textquotedblright. These solutions describe periodic
patterns of period%
\begin{equation}
\lambda=4K(k)\xi\text{, where }K(k)=\int_{0}^{\frac{\pi}{2}}\frac{\mathrm{d}%
t}{\sqrt{1-k^{2}\sin^{2}t}} \label{period}%
\end{equation}
is the complete Jacobian elliptic integral of the first kind. Together with
$k$, it characterizes the segregation, defined as the ratio between the size
of the homogeneous domains, $L=\lambda/2$, and the width of the interface
separating them, $2\xi$. The equation (\ref{period}) and the\ relation
$\xi=\Delta^{-1}$, enable to rewrite this family as :
\begin{equation}
\Psi_{k,\lambda}(x)=\frac{4K(k)\cdot k}{\lambda}\mathrm{Sn}(\frac
{4K(k)}{\lambda}x,k).
\end{equation}

As for $k=1$, $\mathrm{Sn}(x,1)=\tanh(x)$, we recover the usual solution
\begin{equation}
\Psi_{1,\varepsilon}(x)=\frac{\sqrt{\left\vert \varepsilon\right\vert }}%
{2}\tanh(\frac{\sqrt{\left\vert \varepsilon\right\vert }}{2}x).
\end{equation}
associated with a single interface (or soliton)~of width $2/\sqrt{\left\vert
\varepsilon\right\vert }$ that connected the two homogenous phases $\Psi
=\pm\frac{\sqrt{-\varepsilon}}{2}$. Because $K(1)$ diverges, it corresponds to
a strong, or macroscopic, segregation. In the opposite limit (weak segregation
regime), $\Psi_{k,\varepsilon}(x)$ describes a sinusoidal modulation
\begin{equation}
\Psi_{k\rightarrow0,\varepsilon}(x)=k\sqrt{\frac{\left\vert \varepsilon
\right\vert }{2}}\sin(\sqrt{\frac{\left\vert \varepsilon\right\vert }{2}%
}x)=k\frac{2\pi}{\lambda}\sin(\frac{2\pi}{\lambda}x)=kq\sin(qx)
\end{equation}
It will correspond to the Fourier mode\textbf{\ }$q=\frac{2\pi}{\lambda}%
=\frac{\sqrt{\left\vert \varepsilon\right\vert }}{2}$\textbf{\ }of the initial
white noise, with vanishing amplitude\textbf{\ }$\Psi_{\mathbf{q}}%
=kq$\textbf{.} Since experiences, numerical simulations and linear stability
analysis show that $\lambda$, the spatial period of the pattern is constant
during the whole spinodal decomposition process, we choose $\lambda$ to
coincide with the most instable wave length obtain with the Cahn Hilliard
linear approach, $\lambda=\lambda_{C-H}=\frac{4\pi}{\sqrt{-\varepsilon_{0}}}$,
where $\varepsilon_{0}$ is the quench temperature. We then obtain a one
parameter family of profiles $\Psi^{\ast}(x,k)=\Psi_{k,\lambda_{C-H}}(x)$
which describe very well both the linear growth and the saturation of the
spinodal decomposition \cite{pre}. Thus, the dynamics is now reduced to the
time evolution of the single free parameter: $k(t)$.\textbf{\ }Using equations
(\ref{amplitude}) and (\ref{period}), we find that\textbf{\ }$\lambda
$\textbf{, }$k$\textbf{\ }and $\varepsilon$\ are related to one another
through the state equation\textbf{\ }
\begin{equation}
\varepsilon(k)=-2(1+k^{2})\left(  \frac{4K(k)}{\lambda}\right)  ^{2}.
\label{implicit}%
\end{equation}
This implicit equation tells us that if we impose $\lambda=\lambda_{C-H}$, the
dynamics can be reduced to the evolution of $\varepsilon(t)$,\textbf{\ }%
interpreted as the fictitious temperature or \textquotedblleft local
temperature\textquotedblright\ associated with the segregation parameter
$k(t)$ which characterizes the pattern. This temperature can be extracted from
the profile at any time, using the correspondence between $\varepsilon$ and
$k$ of equation (\ref{implicit}). For instance, at $t=0$, the amplitude is
small and we find that $k(t=0)=\frac{\Psi_{\mathbf{q}}\lambda_{C-H}}{2\pi
}\rightarrow0$ and thus $\varepsilon^{\ast}(0)=8\pi^{2}/{\lambda}^{2}$,
different \textit{a priori} from $\varepsilon_{0}$ ($\varepsilon^{\ast
}(0)=\frac{\varepsilon_{0}}{2}$ for $\lambda=\lambda_{C-H}$) : the system is
initially out of equilibrium

Somehow, the dynamics of (C-H) can be projected at first order onto a dynamics
along the sub-family $\Psi^{\ast}(x,k)=\Psi_{k,\lambda_{C-H}}(x) $, which can
be considered as an attractor of the solutions, i.e. the density profile of
the system will evolve with time, staying always close to a function
$\Psi^{\ast}(x,k)$. Thus, using equation (\ref{CH0}), the (C-H) dynamics
reduces to equation :
\begin{equation}
\frac{\partial\Psi^{\ast}}{\partial t}\left(  \mathbf{r},t\right)
=\frac{\partial\Psi^{\ast}}{\partial k}\left(  \mathbf{r},k(t)\right)
\frac{dk}{dt}=\frac{\left(  \varepsilon_{0}-\varepsilon(k)\right)  }%
{2}\mathbf{\nabla}^{2}\Psi^{\ast}. \label{simplified}%
\end{equation}
The non linearity has now disappeared into $\Psi^{\ast}(x,k)$. Using a
solubility condition, it is possible to transform this equation into a
differential equation for $k(t)$ and to compute the full non linear part of
this dynamics (i.e. the saturation of the spinodal decomposition), which leads
the system in a well defined stationary state \cite{pre}.

\subsection{Saturations of the Spinodal Decomposition Dynamics}

The spinodal decomposition dynamics will saturate when the fictitious
temperature $\varepsilon(t)$ will reach the real thermodynamic one, i.e. the
quench temperature $\varepsilon_{0}$; that is, using equation of state
(\ref{implicit}) for $\lambda=\lambda_{C-H}$, when $k=k_{0}^{s}=\!0.687$
solution of the implicit equation :
\begin{equation}
2(1+k_{0}^{s2})K(k_{0}^{s})^{2}=-\frac{\varepsilon_{0}\lambda_{{\small C-H}%
}^{2}}{16}=\pi^{2}\text{ }.
\end{equation}
Using linear stability analysis, Langer has shown that the stationary profile
thus obtained, $\Psi^{\ast}(x,k_{0}^{s})=\Psi_{k_{0}^{s},\lambda_{C-H}}(x)$,
is destroyed by stochastic thermal fluctuations. He has identified the most
instable mode as an \textquotedblright antiferro\textquotedblright\ mode,
leading to a period doubling. The result of this destabilization is another
periodic profile of alternate interfaces of period $\lambda=2\lambda_{C-H}$,
where the length of the domains is : $L=\lambda/2=\lambda_{C-H}$. This means
that the new stationary profile is given by $\Psi_{k_{1}^{s},2\lambda_{C-H}%
}(x)$, where $k_{1}^{s}=\!0.985$ is solution of the implicit relation%
\begin{equation}
2(1+k_{1}^{s2})K(k_{1}^{s})^{2}=\frac{-\varepsilon_{0}(2\lambda_{{\small C-H}%
})^{2}}{16}=4\pi^{2}\text{ }.
\end{equation}
%
%%%%%%%%%%%%%%%%figure 2%%%%%%%%%%%%%%%%%%%%%%%%
%%%%%%%%%%%%%%%%%%%%%%%%%%%%%%%%%%%%%%%%%%%%%%
Again, this new stationary profile turns out to be linearly instable with
respect to an \textquotedblright antiferro\textquotedblright\ perturbation of
period $4\lambda_{C-H}$.

Thus these families of profiles and instabilities enable to describe the one
dimensional coarsening as a cascade of doubling process. Each of these
successive intermediate profiles can be described by an element of the above
family of soliton lattice $\Psi_{k_{n}^{s},2^{n}\times\lambda_{C-H}}(x)$. The
sequence of associated segregation parameters $\left\{  k_{n}^{s}\right\}  $,
are determined by the implicit relations
\begin{equation}
2(1+k_{n}^{s2})K(k_{n}^{s})^{2}=-\frac{\varepsilon_{0}(2^{n}\lambda
_{{\small C-H}})^{2}}{16}=\pi^{2}2^{2n}. \label{implicitn}%
\end{equation}
We have found numerically for the first of them \cite{universality}
\begin{equation}%
\begin{tabular}
[c]{|l|}\hline
$k_{0}^{s}\!=\!0.6869795924$\\\hline
$k_{1}^{s}\!=\!0.9851675587$\\\hline
$k_{2}^{s}\!=0.99997210165$\\\hline
$k_{3}^{s}\!=\!0.9999999999027$\\\hline
\end{tabular}
\
\end{equation}
We see that $\left\{  k_{n}^{s}\right\}  $ converges toward $k_{\infty}^{s}=1$
(single interface case), while the amplitude of the modulation converges
toward $\frac{\sqrt{\left\vert \varepsilon_{0}\right\vert }}{2}$. For large
$n$, we can conclude from the implicit relation (\ref{implicitn}) that the
ratio of the domain size to the interface width characterized by $K(k_{n}%
^{s})$ behaves as $\pi2^{n-1}$ and $(k_{n}^{s})^{2}=1-16\exp(-\pi2^{n})$. Each
of the stationary profiles
\begin{equation}
\Psi_{n}(x)=\Psi_{k_{n}^{s},2^{n}\lambda_{{\small C-H}}}(x)=\frac
{\sqrt{\left\vert \varepsilon_{0}\right\vert }k_{n}^{s}K(k_{n}^{s})}{2^{n}\pi
}\mathrm{Sn}(\frac{\sqrt{\left\vert \varepsilon_{0}\right\vert }K(k_{n}^{s}%
)}{2^{n}\pi}x,k_{n}^{s})=\frac{\sqrt{\left\vert \varepsilon_{0}\right\vert
}k_{n}^{s}}{\sqrt{2(1+k_{n}^{s2})}}\mathrm{Sn}(\frac{\sqrt{\left\vert
\varepsilon_{0}\right\vert }x}{\sqrt{2(1+k_{n}^{s2})}},k_{n}^{s})
\label{ansatzn}%
\end{equation}
is identically destroyed by the Langer \textquotedblright
antiferro\textquotedblright\ instability .

\section{1D Ostwald Ripening}

\subsection{Non-symmetric soliton lattice profile as an ansatz for the 1D
coarsening process}

In order to describe one step of the coalescence process, i.e. the dynamics
that starts from $\Psi_{n}(x)$ and ends with the profile $\Psi_{n+1}(x)$ , we
will use another family of equilibrium profiles \cite{novik}, of period
$2\lambda$, solutions of (G-L) equation, which write:%
\begin{equation}
\widehat{\psi}(x,k)=\frac{K(k)}{\lambda}+\widehat{\psi}^{\ast}(x,k)=\frac
{K(k)}{\lambda}\frac{1-k^{\prime}-(1+k^{\prime})\sqrt{1-k^{\prime}}%
Sn(2x,k)}{\sqrt{1-k^{\prime}}Sn(2x,k)-1}%
\end{equation}
where $k^{2}+k^{\prime2}=1$. Note that $<\widehat{\psi}(x,k)>=0$ whereas the
family $\widehat{\psi}^{\ast}(x,k)$ is solution of :%
\begin{equation}
\nabla^{2}\widehat{\psi}^{\ast}=\varepsilon(k)\frac{\widehat{\psi}^{\ast}}%
{2}+2\widehat{\psi}^{\ast3}+\mu(k) \label{eigen}%
\end{equation}
with $\mu(a,k)=\left(  k^{2}-1\right)  \left(  \frac{2K(k)}{\lambda}\right)
^{3}$ and $\varepsilon(k)=(k^{2}-5)\left(  \frac{2K(k)}{\lambda}\right)  ^{2}$
\ (we have $\varepsilon(k)>\varepsilon_{0}$ ).

Using Gauss' transformation (or descending Landen transformation
\cite{abra}),\ we can relate the soliton lattice of spatial period $2\lambda$
(and of modulus $k$) to the soliton lattice of period $\lambda$ (and of
modulus $\mu=\frac{1-k^{\prime}}{1+k^{\prime}}$) as follows%
\begin{align}
&
\begin{array}
[c]{ccc}%
\widehat{\psi}(x-\frac{\lambda}{2},k)+\widehat{\psi}(x+\frac{\lambda}{2},k) &
= & \frac{4\mu K(\mu)}{\lambda}Sn((4x+\lambda)\frac{K(\mu)}{\lambda},\mu)
\end{array}
\label{initial}\\
&
\begin{array}
[c]{ccc}%
\widehat{\psi}(x-\frac{\lambda}{4},k)+\widehat{\psi}(x+\frac{\lambda}{4},k) &
= & \frac{4kK(k)}{2\lambda}Sn(4x\frac{K(k)}{2\lambda},k)
\end{array}
\label{final}%
\end{align}

Note that $K(k)=\left(  1+\mu\right)  K(\mu)$. As if $k=k_{n+1}^{s}$,
$\mu=k_{n}^{s}$, we see that both the initial state $\Psi_{n}(x)=\Psi
_{k_{n}^{s},2^{n}\lambda_{C-H}}(x)$ and the final state $\Psi_{n+1}%
(x)=\Psi_{k_{n+1}^{s},2^{n+1}\lambda_{C-H}}(x)$ of a step of the coalescence
process can be describe, modulo a phase shift, by the same function :
\begin{equation}
\Phi(x,k,\phi)=\widehat{\psi}(x-(1-\phi/2)\frac{\lambda}{2},k)+\widehat{\psi
}(x+(1-\phi/2)\frac{\lambda}{2},k) \label{magic}%
\end{equation}
with $k=k_{n+1}^{s}$ and $\lambda=2^{n}\lambda_{C-H}$ (see also Figure 1).
Therefore we can describe the coalescence by a transformation at constant
segregation parameter $k$, while the degree of freedom $\phi$, associated with
the relative phase between the two profiles, evolves in time from $0$ to $1$
(or $-1$) according to the C-H dynamics \cite{annexe}.

\subsection{Energy Landscape}

In order to prove the usefulness of this ansatz, we have plot the energy
averaged over a period, $\mathcal{F}_{GL}(\phi)=\int F_{GL}(\Phi
(x,k,\phi))\mathrm{d}x$, as a function of the parameter $\phi$, keeping $k$
constant. We see in Figure 2 that the value $\phi=0$ corresponds to a local
maximum of energy, while $\phi=1$ (or $-1$) is a minimum. Note that there is
no energy barrier in this particular energy landscape, in agreement with
linear stability analysis.
%
%%%%%%%%%%%%%%%%figure 5%%%%%%%%%%%%%%%%%%%%%%%%
%%%%%%%%%%%%%%%%%%%%%%%%%%%%%%%%%%%%%%%%%%%%%%

Using the properties of the ansatz, we can extract $\phi(t)$ from a numerical
simulation. Indeed
\begin{align}
\Phi(\frac{\lambda}{2},k,\phi)  &  =\widehat{\psi}(\phi\frac{\lambda}%
{4},k)+\widehat{\psi}(\lambda-\phi\frac{\lambda}{4},k)=2\widehat{\psi}%
(\phi\frac{\lambda}{4},k)\\
\Phi(-\frac{\lambda}{2},k,\phi)  &  =\widehat{\psi}(\phi\frac{\lambda}%
{4}-\lambda,k)+\widehat{\psi}(-\phi\frac{\lambda}{4},k)=2\widehat{\psi}%
(-\phi\frac{\lambda}{4},k)\nonumber
\end{align}
For $\phi=0$, $\Phi(\frac{\lambda}{2})=\Phi(-\frac{\lambda}{2})=\frac
{2K}{\lambda}\left(  k^{\prime}-1\right)  $ while $\Phi(\frac{\lambda}%
{2})=-\Phi(-\frac{\lambda}{2})=\frac{2Kk}{\lambda}$ for $\phi=1$. A
$k=k_{n+1}^{s}$ is known, from $A(t)\in\lbrack0,1]$ defined as follows%
\begin{equation}
2A(t)=1-\frac{\Phi(\frac{\lambda}{2},k,\phi(t))}{\Phi(-\frac{\lambda}%
{2},k,\phi(t))}=\frac{4k^{\prime}Sn(\phi\frac{K(k)}{2},k)}{\sqrt{1-k^{\prime}%
}+2k^{\prime}Sn(\phi\frac{K(k)}{2},k)-k\sqrt{1+k^{\prime}}Sn^{2}(\phi
\frac{K(k)}{2},k)},
\end{equation}
one can extract $\phi$ :
\begin{equation}
\phi(t)=\frac{2}{K(k)}\int_{0}^{\frac{k^{\prime}(A-1)+\sqrt{A^{2}+k^{\prime
2}-2Ak^{\prime2}}}{kA\sqrt{1+k^{\prime}}}}\frac{da}{\sqrt{1-k^{2}a^{2}}%
\sqrt{1-a^{2}}}. \label{phase}%
\end{equation}

One sees in Figure3 that the dynamics is dominated by the time required to
leave a stationary state. Moreover, the dynamics ends when $\phi$ reaches 1~:
we then have $\Phi=\Psi_{n+1}$ for which $\frac{\varepsilon_{0}}{2}\Psi
_{n+1}+2\Psi_{n+1}^{3}-\nabla^{2}\Psi_{n+1}=0$.

\subsection{Linear stability analysis}

If we look at the time evolution of the profile $\Phi(x,k_{n+1}^{s},\phi)$,
starting from the region $\phi=0$, we can transform the (C-H) equation into a
phase field equation, replacing $\frac{\partial}{\partial t}$ $\Phi
(x,k_{n+1}^{s},\phi)$ \ by $\frac{\partial}{\partial\phi}\Phi(x,k_{n+1}%
^{s},\phi(t))\times\frac{d\phi}{dt}$ (with $k$ fixed). The dynamics will then
be similar to spinodal decomposition (eq. (\ref{simplified})), for $\phi$
growing initially as $\exp(t/\tau_{n})$ and saturating later at $\phi=1$.

In Figure 4 is plotted $\frac{\partial\Phi}{\partial\phi}(x,k_{n+1}^{s}%
)=\frac{\lambda}{4}\widetilde{\Psi}_{L}(x,k_{n+1}^{s})$ for $\phi=0$, which
corresponds to the most unstable mode founded in Langer's linear stability
analysis and is characterized by the alternated growth and decrease of domains
(\textquotedblright antiferro\textquotedblright\ mode) :%
\begin{equation}
\widetilde{\Psi}_{L}(x,k)=\widehat{\psi}^{\prime}(x-(1-\phi/2)\frac{\lambda
}{2},k)-\widehat{\psi}^{\prime}(x+(1-\phi/2)\frac{\lambda}{2},k)\text{ }%
\end{equation}
where%
\begin{equation}
\widehat{\psi}^{\prime}(x,k)=k^{\prime}\sqrt{1-k^{\prime}}\left(  \frac
{2K(k)}{\lambda}\right)  ^{2}\frac{Cn(2x\frac{K(k)}{\lambda},k)Dn(2x\frac
{K(k)}{\lambda},k)}{\left(  1-\sqrt{1-k^{\prime}}Sn(2x\frac{K(k)}{\lambda
},k)\right)  ^{2}}%
\end{equation}
verifies
\begin{equation}
\nabla^{2}\widehat{\psi}^{\prime}(x,k)=\varepsilon(k)\frac{\widehat{\psi
}^{\prime}(k,x)}{2}+6\widehat{\psi}^{\ast2}\widehat{\psi}^{\prime}(k,x).
\label{semianzatz}%
\end{equation}

In order to describe the evolution of the phase $\phi(t)$ we linearize the
Cahn-Hilliard dynamics around $\Psi_{n}(x)=\Phi(x,k_{n+1}^{s},\phi
=0)\ $inserting $\Psi\left(  x,t\right)  =\Psi_{n}(x)+\frac{\partial\Phi
}{\partial\phi}\phi(t)=$ $\Psi_{n}(x)+\frac{\lambda}{4}\widetilde{\Psi}%
_{L}(x,k_{n+1}^{s})\phi(t)$ into (\ref{CHeq}). We then have the following dynamics%

\begin{equation}
\widetilde{\Psi}_{L}.\frac{\mathrm{d}\phi}{\mathrm{d}t}=\phi(t)\frac
{\partial^{2}}{\partial x^{2}}\left(  \frac{\varepsilon_{0}}{2}\widetilde
{\Psi}_{L}+6\Psi_{n}^{2}\widetilde{\Psi}_{L}-\nabla^{2}\widetilde{\Psi}%
_{L}\right)  =\phi(t)\frac{\partial^{2}}{\partial x^{2}}\mathcal{L}%
(\widetilde{\Psi}_{L}). \label{lame}%
\end{equation}
where $\mathcal{L}$ is the Lam\'{e} operator. Even if this operator doesn't
have simple (algebraic) exact eigenfunction of period $2\lambda_{C-H}$
\cite{arscott}, $\widetilde{\Psi}_{L}(x,k,\phi)$, for $\phi=0$ and
$k=k_{n+1}^{s}$, happens nevertheless to be a good approximation for the
eigenfunction of lowest eigenvalue \cite{annexe}. Due to the concavity of
$\mathcal{F}_{GL}(\phi)$ around $\phi=0$, (see Figure 2), this eigenvalue will
be negative, triggering a linear destabilization of the pattern $\Psi_{n}$ and
an exponential amplification of the perturbation,\ i.e. an exponential growth
of the translation $\phi$\ with time.

Indeed, using equation (\ref{eigen}) we get \
\begin{equation}
\nabla^{2}\widetilde{\Psi}_{L}=\nabla^{2}\left[  \widehat{\psi}_{+}^{\prime
}-\widehat{\psi}_{-}^{\prime}\right]  =\frac{\varepsilon(k)}{2}\widetilde
{\Psi}_{L}+6(\widehat{\psi}_{+}^{\ast2}\widehat{\psi}_{+}^{\prime}%
-\widehat{\psi}_{-}^{\ast2}\widehat{\psi}_{-}^{\prime}).
\end{equation}

So%
\begin{align}
\frac{\varepsilon_{0}}{2}\widetilde{\Psi}_{L}+6\Psi_{i}^{\ast2}\widetilde
{\Psi}_{L}-\nabla^{2}\widetilde{\Psi}_{L}  &  =\frac{\varepsilon
_{0}-\varepsilon(k)}{2}\widetilde{\Psi}_{L}+6\left(  \widehat{\psi}_{-}%
^{\ast2}\widehat{\psi}_{+}^{\prime}-\widehat{\psi}_{+}^{\ast2}\widehat{\psi
}_{-}^{\prime}\right) \\
&  +6\left[  (\frac{2K(k)}{\lambda})^{2}+2(\frac{2K(k)}{\lambda}(\widehat
{\psi}_{+}^{\ast}+\widehat{\psi}_{-}^{\ast})+\widehat{\psi}_{+}^{\ast}%
\widehat{\psi}_{-}^{\ast})\right]  (\widehat{\psi}_{+}^{\prime}-\widehat{\psi
}_{-}^{\prime}).\nonumber
\end{align}

It turns out that for $\phi=0$,
\[
\frac{2K(k)}{\lambda}\left(  \widehat{\psi}_{+}^{\ast}+\widehat{\psi}%
_{-}^{\ast}\right)  +\widehat{\psi}_{+}^{\ast}\widehat{\psi}_{-}^{\ast
}=\left(  k^{\prime2}-4\right)  \left(  \frac{K(k)}{\lambda}\right)  ^{2}%
\]
together with
\[
\widehat{\psi}_{-}^{\ast2}\widehat{\psi}_{+}^{\prime}-\widehat{\psi}_{+}%
^{\ast2}\widehat{\psi}_{-}^{\prime}\simeq\left(  4-k^{\prime2}-4k^{^{\prime}%
}\right)  \left(  \frac{K(k)}{\lambda}\right)  ^{2}(\widehat{\psi}_{+}%
^{\prime}-\widehat{\psi}_{-}^{\prime}).
\]
So finally\
\begin{equation}
\frac{\varepsilon_{0}}{2}\widetilde{\Psi}_{L}+6\Psi_{i0}^{\ast2}%
\widetilde{\Psi}_{L}-\nabla^{2}\widetilde{\Psi}_{L}\simeq-6k^{^{\prime}%
}\left(  \frac{2K(k)}{\lambda}\right)  ^{2}\widetilde{\Psi}_{L}.
\end{equation}
As the period associated with $k_{n+1}^{s}$ is $2\lambda,$ equation
(\ref{implicit}) gives $\left(  \frac{2K(k)}{\lambda}\right)  ^{2}%
=-\frac{\varepsilon_{0}}{2(1+k^{2})}$ and thus equation (\ref{lame})\ can now
be written in a simpler form, similar to equation (\ref{simplified}) :
\begin{equation}
\widetilde{\Psi}_{L}.\frac{\mathrm{d}\phi}{\mathrm{d}t}=\varepsilon_{0}%
\frac{3k^{\prime}}{1+k^{2}}\phi(t)\frac{\partial^{2}}{\partial x^{2}%
}\widetilde{\Psi}_{L}.
\end{equation}
We recover that $\widetilde{\Psi}_{L}$ is an eigenstate of Lam\'{e} equation
with a negative eigenvalue $\frac{-3k^{\prime}\left\vert \varepsilon
_{0}\right\vert }{2-k^{\prime2}}$. This eigenvalue goes to zero as the
coalescence progresses towards higher segregation.

The characteristic time for one step of period doubling is thus $\tau
_{n}\simeq\frac{2}{3\varepsilon_{0}^{2}}\frac{2-k^{\prime2}}{k^{\prime}}$ (the
extra $\varepsilon_{0}/2$ comes from $\frac{\partial^{2}}{\partial x^{2}%
}\widetilde{\Psi}_{L}$). As for small $k^{\prime}$, $k^{\prime}\simeq
4\exp(-K(k))$, we have the following relationship between $\tau_{n}^{-1}$ and
the period of the profile :
\begin{equation}
\tau_{n}^{-1}\simeq\frac{3}{4}\varepsilon_{0}^{2}k^{\prime}=3\varepsilon
_{0}^{2}\exp(-K(k))=3\varepsilon_{0}^{2}\exp(-2^{n-1}\pi)=3\varepsilon_{0}%
^{2}\exp(-\frac{\pi\lambda}{2\lambda_{C-H}}).
\end{equation}
As $\ln\tau_{n}$ goes like a power of $n$, we can conclude that in $D=1$, the
size of the domain will evolve for long $t$ as $\frac{2}{\pi}\lambda_{C-H}%
\ln(3\varepsilon_{0}^{2}t).$

\section{Application : Modulated phase systems}

We can use the preceding ansatz to work out the period of modulated phase
systems for which there is a competition between two types of interactions: a
short-range interaction which tends to make the system more homogeneous
together with a long-range one, or a non-local one, which prefers
proliferation of domain walls. This competition results in a microphase
separation with a preferred length scale \cite{SC}. These systems can be study
using a modified Landau-Ginzburg approach, derived from Cahn-Hilliard
equation~and often use for numerical simulations \cite{brasil}:%
\begin{equation}
{\frac{\partial\Psi}{\partial t}}=(\nabla^{2}\frac{\delta F_{GL}(\Psi)}%
{\delta\Psi})-\beta^{2}\Psi=\mathbf{\nabla}^{2}(\frac{\varepsilon_{0}}{2}%
\Psi+2\Psi^{3}-\mathbf{\nabla}^{2}\Psi)-\beta^{2}\Psi. \label{oono}%
\end{equation}

$-\beta^{2}\Psi$ models in the Cahn-Hilliard equation the long-range
interactions, which prevents the formation of macroscopic domains and favors
the modulation. We could have chosen other ways of representing this
long-range interaction, but the inclusion of such a term, following Oono,
enables to describe the behavior of modulated systems at $T$ much lower than
$T_{c}$, as we will show below. If we suppose for example that in a 3D
problem, the long range interaction decreases like $\frac{1}{r}$, the full
free energy density writes
\begin{equation}
F(\Psi)=F_{GL}+F_{int}=\frac{1}{2}(\nabla\Psi(r))^{2}+\frac{\varepsilon}%
{4}\Psi^{2}(r)+\frac{1}{2}\Psi^{4}(r)+\int\Psi(r^{\prime})g(r^{\prime}%
,r)\Psi(r)\mathrm{d}r^{\prime}\text{ ,}%
\end{equation}

where $g(r^{\prime},r)=\frac{\beta^{2}}{\left\vert r^{\prime}-r\right\vert }%
.$The long range interaction $g(r^{\prime},r)$ corresponds to a repulsive
interaction when $\Psi(r^{\prime})$ and $\Psi(r)$ are of the same sign : thus
it \ favor the formation of interphases. If we want to study the dynamic of
this phase separation, we use the Cahn-Hilliard equation~:
\begin{equation}
{\frac{\partial\Psi}{\partial t}}=\nabla_{r}^{2}\left(  \frac{\delta F(\Psi
)}{\delta\Psi}\right)  =\nabla_{r}^{2}\left(  \frac{\varepsilon_{0}}{2}%
\Psi+2\Psi^{3}-\mathbf{\nabla}^{2}\Psi+\int\Psi(r^{\prime})g(r^{\prime
},r)\mathrm{d}r^{\prime}\right)  .
\end{equation}
If one recalls that $\frac{-1}{\left\vert r^{\prime}-r\right\vert }$ is the
Green's function associated with the Laplacian operator $\nabla_{r}^{2}$ in
3D, the preceding equation then transforms into
\begin{equation}
\nabla_{r}^{2}\left(  \int\Psi(r^{\prime})g(r^{\prime},r)\mathrm{d}r^{\prime
}\right)  =\int\Psi(r^{\prime})\nabla_{r}^{2}g(r^{\prime},r)\mathrm{d}%
r^{\prime}=-\beta^{2}\int\Psi(r^{\prime})\delta(r^{\prime},r)\mathrm{d}%
r^{\prime}=-\beta^{2}\Psi(r).
\end{equation}
which leads to equation (\ref{oono}). The family (\ref{ansatzn}) is not
anymore an exact stationary solution of this dynamics. Nevertheless, $\Psi
_{n}(x,k_{n},2^{n}\lambda_{{\small C-H}})$ is an approximate solution and thus
can be use as a tool for the calculation using a solubility condition (i.e. if
we project the dynamics on $\chi\in Ker(\frac{\partial^{2}}{\partial x^{2}%
}\mathcal{L)}$). There exists a new sequence $\{k_{n}^{o}\}$ which satisfies
$(\varepsilon^{\ast}(k_{n}^{o})-\varepsilon_{0})<\chi\mid\partial_{x^{2}}%
\Psi_{k_{n}^{o}}>+\beta^{2}<\chi\mid\Psi_{k_{n}^{o}}>=0$ and which will
characterize the new family of approximate stationary configuration. When
looking at the linear stability analysis of these solutions, equation
(\ref{CHeq}) now writes%
\begin{equation}
\widetilde{\Psi}_{L}.\frac{\mathrm{d}\phi}{\mathrm{d}t}=\phi(t)\left(
\frac{3\varepsilon_{0}k^{\prime}}{2-k^{\prime2}}\frac{\partial^{2}}{\partial
x^{2}}\widetilde{\Psi}_{L}-\beta^{2}\widetilde{\Psi}_{L}\right)  .
\end{equation}
The profile will become stable when the eigenvalue of this modified Lam\'{e}'s
equation become negative, i.e. for $\tau_{n}^{-1}<\beta^{2}$. This will take
place for $3\varepsilon_{0}^{2}\exp(-\frac{\pi\lambda}{2\lambda_{C-H}}%
)\simeq\beta^{2}$. The dynamics of period doubling will thus end with a
thermodynamically stable 1D microseparated phase of period $\lambda=\frac
{8}{\sqrt{-\varepsilon_{0}}}\ln(\frac{3\varepsilon_{0}^{2}}{\beta^{2}})$

\section{Discussion on the Hypothesis and Conclusion}

We have shown that the choice of two ansatzs within the soliton-lattice family
allows a reliable description of the one dimensional dynamics of both spinodal
decomposition and Ostwald ripening. Contrary to \cite{baron}, our ansatz
relies on the hypothesis that during the first stage of the dynamics, the
periodicity of the order parameter remains constant, while during each steps
of the coarsening process, it is the parameter $k$ which remains constant. In
a sense, there are adiabatic ansatzs : the generation of higher harmonics is
governed solely by the time evolution of $k(t)$ during the non linear
saturation of the spinodal decomposition and later on solely by $\phi(t)$
during the Ostwald ripening. The validity of the assumptions have been
investigated in details and checked numerically \cite{pre}.

Our analytic method rely on the assumption that at each\ step of the dynamics,
the system can be characterized by a specific spatial period : we need
therefore to discuss how this approach is relevant to the general case where
noise is present. We have noted numerically that, for the spinodal
decomposition, the average size of the modulation is $\lambda_{C-H}$, with a
deviation of less than one percent from the value predicted by the linear
theory. It does not mean that, in a real system, each domain has a length
scale of $L=\lambda_{C-H}/2$, but that the distribution of the domains' length
will be centered around $L$. The coalescence\ events can be neglected during
the initial growth of the\ amplitude of the modulation : as all the
eigenvalues of the Lam\'{e} operator (\ref{lame}) are then positive, the
dynamics remains within the ansatz subfamily $\Psi^{\ast}(x,k)$. Only after
this initial growth has saturated, turns negative the lowest eigenvalue ; the
coalescence process then starts, and dominates the forthcoming dynamics.

We have computed in this article the characteristic time associated with one
step of ideal coalescence. By ideal coalescence, we mean a process which
breaks as few symmetries as possible. In a real system, because of\ initial
fluctuations in the periodicity of the pattern selected just after the quench,
this instability will concern only region of finite size, where it choose a
certain sublattice, or a range for $\phi$ (for example, $\phi$ varies from $0$
to $1$), while it is the opposite choice in the neighboring region ( $\phi$
varies from $0$ to -$1$). The global symmetry is thus recover on the overall
(as in an antiferromagnet). During each step of the process, the width of the
domains will locally double; but for the system as a whole, due to non
synchronization between regions, the average length scale will vary continuously.

We have also shown that in a modified version of the Cahn-Hilliard dynamics,
which take into account long range interactions, the computation of this
lowest eigenvalue enables to compute that the spatial period of the
thermodynamically stable modulated phase will be $\lambda=\frac{8}%
{\sqrt{-\varepsilon_{0}}}\ln(\frac{\varepsilon_{0}^{2}}{2\beta^{2}}).$

The use of the solubility technics combined with the choice of an adiabatic
ansatz might be generalized to the study of other non linear dynamics. For
instance, spinodal decomposition in superfluid Helium or Bose condensate has
been argued to be described by a cubic-quintic non linear equation\cite{rica}%
~; in this particular case, one needs to retrieve a relevant soliton-like
family of solution along which to compute the adiabatic dynamics. The same
difficulties would arise as well when the method will be adapted to higher
space dimensions as there is no equivalent to soliton lattice in 2D; the
ansatzs (\ref{ansatzn}) and (\ref{magic}) could however be used in a numerical
simulation as tools for following the dynamics.

It might also be possible to extend this approach to non symmetric profile
like the ones introduced in different contexts\cite{novik,yi}. Or for the
cases of special quenches which are time periodic \cite{onuki} or spatially
periodic \cite{delville}.

I would like to thanks Christophe Josserand, Avadh Saxena and David Andelman
for fruitful discussions and Turab Lookman for bringing to me ref
\cite{brasil}.

\newpage

\section{Figures}

Figure 1

{\small Construction of the two first steady solutions of the (C-H)
dynamics},{\small \ using a superposition of the non-symmetric profile
}$\widehat{\psi}^{\ast}(k,x)${\small , itself stationary solution of the (C-H)
equation. By changing the phase shift between the two profiles entering into
the linear combination, one obtains two different symmetric profiles, of
periods }$\lambda${\small \ and segregation parameter }$k_{0}^{s}%
\!=\!0.687$\ {\small (equation (\ref{initial}))\ or of period }$2\lambda$
{\small and\ segregation parameter }$k_{1}^{s}\!=\!0.985$ {\small (equation
(\ref{final}))}.

\bigskip

Figure 2

{\small Profile of the free energy landscape during a coarsening
process,\ F(}$\phi${\small ). It starts at }$\phi=0${\small \ for a
configuration characterized by the segregation ratio }$k_{1}^{s}%
\!=\!0.687${\small \ for which the energy per unit length is F(}$\phi
${\small )}$\simeq-0.135${\small ; one sees\ that in this region, the free
energy is a concave function of $\phi$\ and thus, the associated pattern is
linearly instable. The elementary step of the coarsening process ends for
}$\phi=1${\small \ associated with a pattern characterized by the segregation
ratio }$k_{2}^{s}\!=\!0.985${\small \ for which the energy per unit length is
F(}$\phi${\small )}$\simeq-0.45$.{\small \ In the region }$\phi=1,$%
{\small \ the free energy is a convex function of }$\phi${\small . Note that
there is no energy barrier}.

\bigskip

Figure 3

{\small Parameter }$\phi${\small \ as a function of time. }$\phi${\small \ is
extract from a numerical integration of the Cahn-Hilliard dynamics using
relation (\ref{phase}) to relate at a given time the profile with the phase
}$\phi${\small . It starts at }$\phi=0${\small \ with an exponential growth
and saturates at }$\phi=1${\small .}

\bigskip

Figure 4

{\small Langer's most instable perturbation mode of destabilization of the
soliton lattice is identified with }$\widetilde{\Psi}_{L}=\frac{\partial
}{\partial\phi}${\small \ }$\Phi(x,k,\phi)${\small \ at }$\phi=0.${\small \ It
is composed of two antisymmetric patterns, plotted in dotted (plain) line,
evolving toward right (left) at velocity +}$\frac{d\phi}{dt}$ {\small (-}%
$\frac{d\phi}{dt}${\small ), causing an \textquotedblright
antiferro\textquotedblright\ instability leading to a period doubling of the
pattern. They are the spatial derivative of the initial non symmetric profile
}$\widehat{\psi}^{\ast}(x)${\small \ which has been used to construct our
ansatz in Figure 1.}


\begin{thebibliography}{99}
\bibitem {universality}Note that if we had looked at the equation
$\frac{\partial\Psi}{\partial t}=\mathbf{\nabla}^{2}(\alpha\Psi+\beta\Psi
^{3}-\mathbf{\nabla}^{2}\Psi)$, the amplification factor would have been
$\sigma(q)=-(q^{2}+\alpha)q^{2}$ for which the most unstable mode is
$\lambda_{C-H}=2\sqrt{2}\pi/\sqrt{-\alpha}$. Meanwhile, the period of the
static Ginzburg-Landau solution (\ref{amplitude})\ would be $\xi=2\beta
^{-1}\Delta^{-1}=\sqrt{\frac{k^{2}+1}{-\alpha}}$leading to the same implicit
relation (\ref{implicit}) and same sequence $\{k_{n}\}$.
\end{thebibliography}
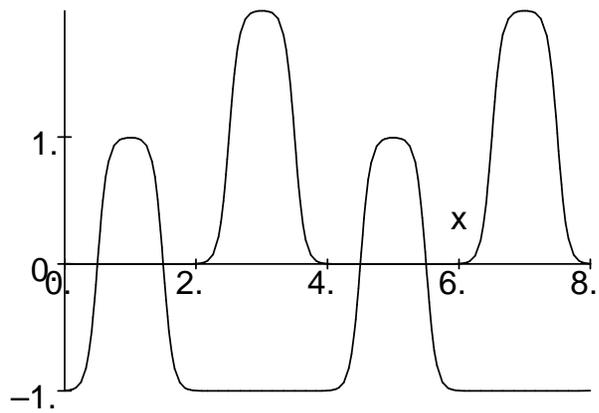
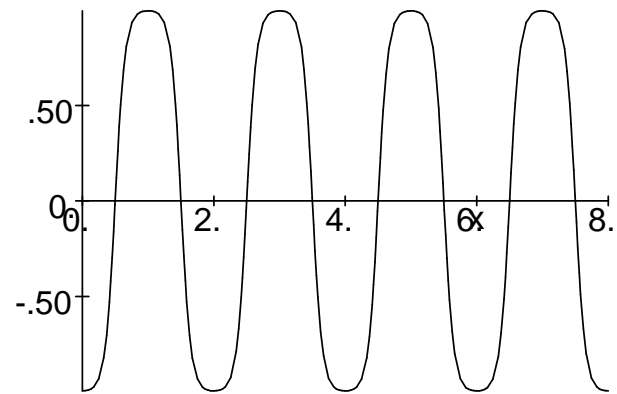
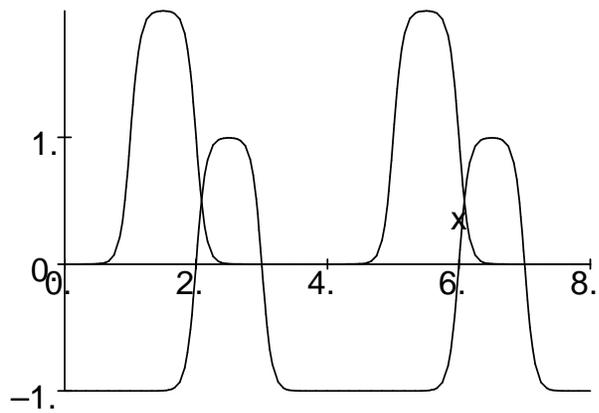
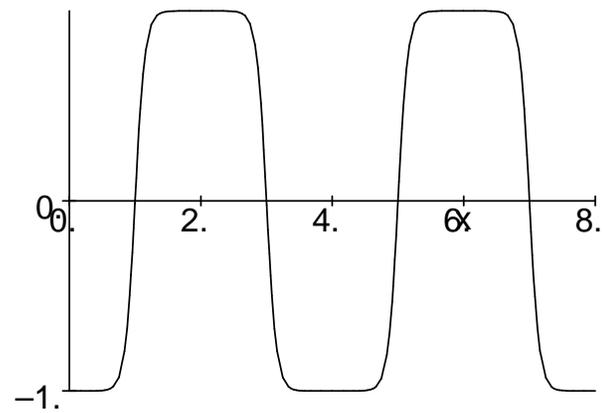
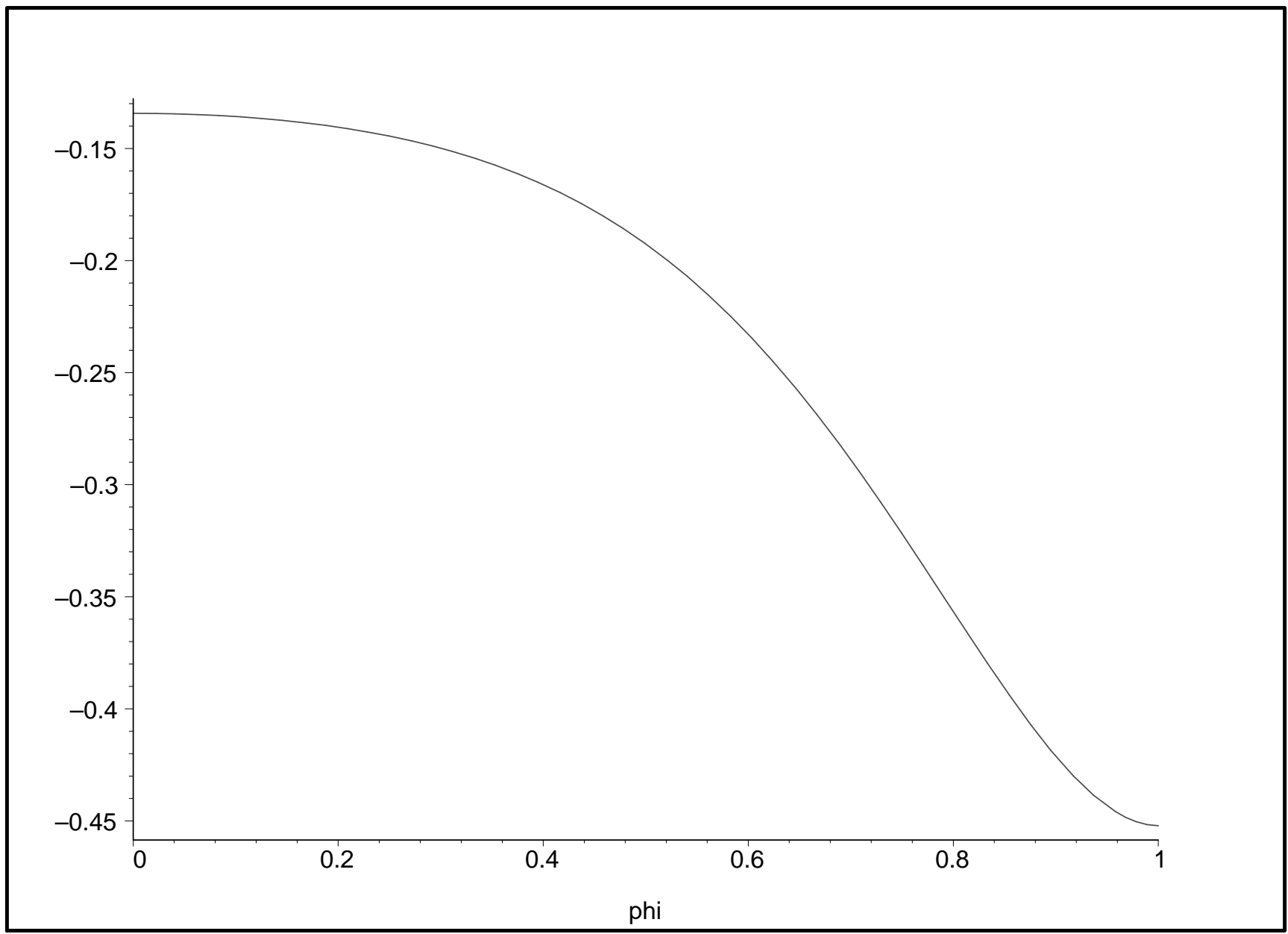
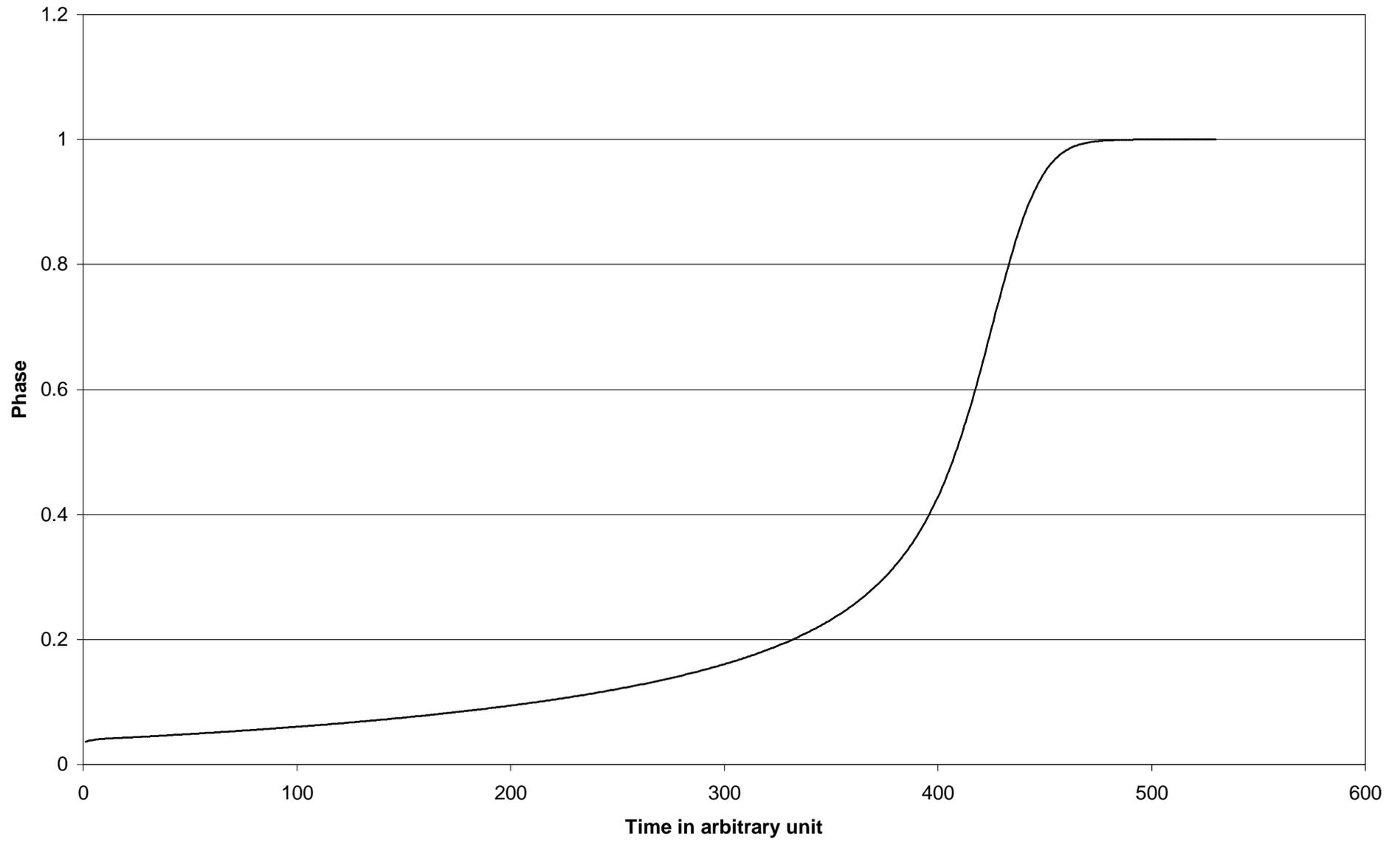
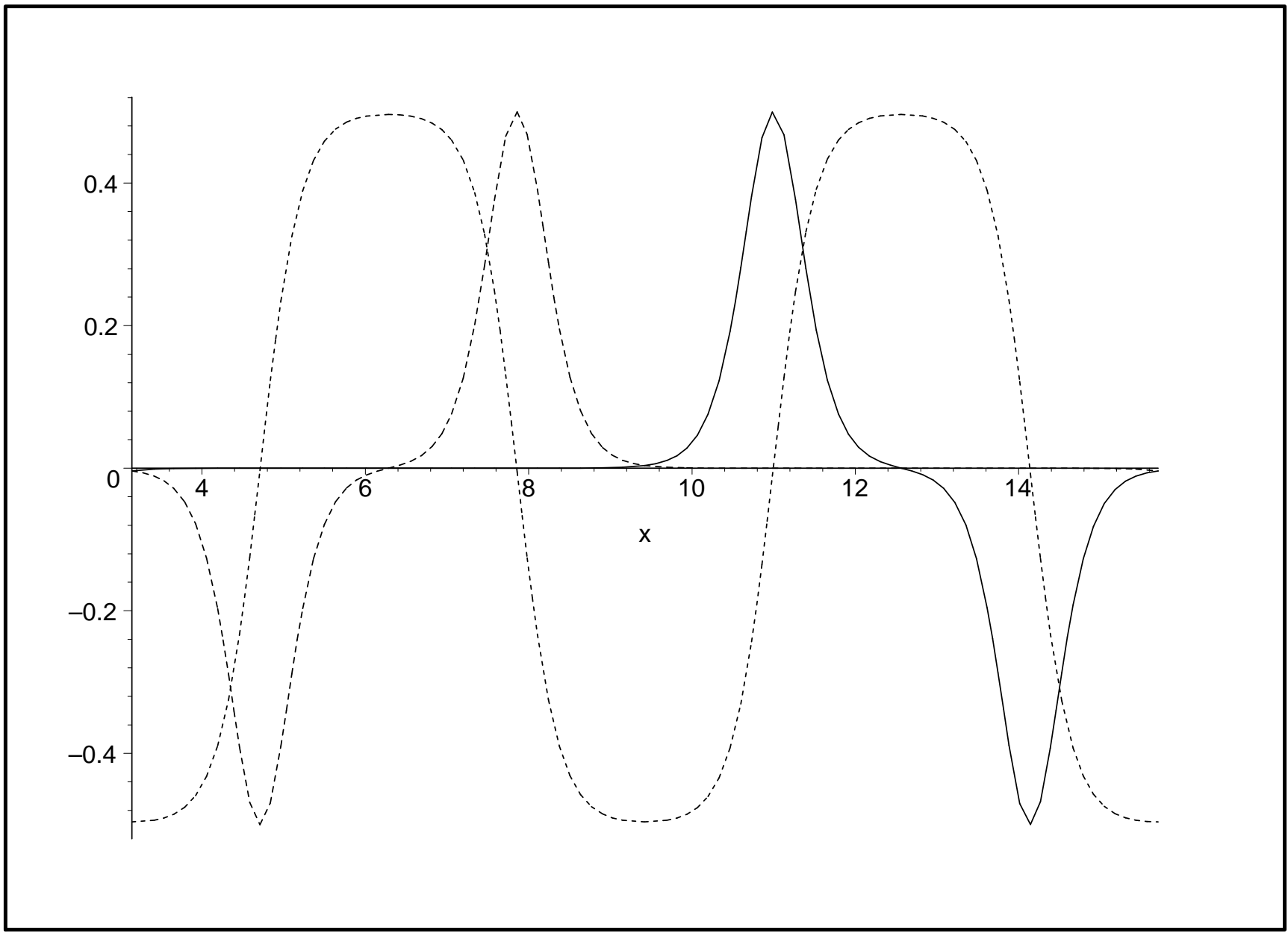
\end{document}